\documentclass{svjour3}                     % onecolumn (standard format)
\smartqed  % flush right qed marks, e.g. at end of proof
\usepackage{graphicx}
\newcommand{\bm}[1]{ \mbox{\boldmath $#1$}  }

\begin{document}

\title{Relativistic description of $^3$He($e,e^\prime p$)$^2$H
  \footnote{Presented at the 21st European Conference on Few-Body
    Problems in Physics, Salamanca, Spain, 30 August - 3 September
    2010.}  }


\author{R.~\'Alvarez-Rodr\'{\i}guez
\and
J.M.~Ud\'{\i}as
\and
J.R.~Vignote
\and
E.~Garrido
\and
P.~Sarriguren
\and
E.~Moya~de~Guerra
\and
E.~Pace
\and
A.~Kievsky
\and
G.~Salm\`e 
}


\institute{R. \'Alvarez-Rodr\'{\i}guez, J.M. Ud\'{\i}as and E. Moya de
  Guerra \at Grupo de F\'{\i}sica Nuclear, Departamento de F\'{\i}sica
  At\'omica, Molecular y Nuclear, Universidad Complutense de Madrid \\ E-28040 Madrid (Spain)\\
  \email{jose@nuc2.fis.ucm.es (J.M. Ud\'{\i}as)} % \\
  \and J.R. Vignote, E. Garrido and P. Sarriguren \at Instituto de
  Estructura de la Materia, Consejo Superior de Investigaciones
  Cient\'{\i}ficas \\ E-28006 Madrid (Spain) 
\and E. Pace \at Dipartimento di Fisica, Universit\`a degli Studi di Roma ``Tor Vergata'', and INFN \\ I-00133 Rome (Italy)
\and A. Kievsky \at Dipartimento di Fisica, Universit\`a di Pisa, and INFN \\ I-56127 Pisa (Italy) 
\and G. Salm\`e \at Istituto Nazionale di Fisica Nucleare, Sezione di Roma \\ I-00185 Rome (Italy)}

\date{Received: date / Accepted: date}

\maketitle

\begin{abstract}
  The Relativistic Distorted-Wave Impulse Approximation is used to
  describe the $^3$He($e,e^\prime p$)$^2$H process.  We describe the
  $^3$He nucleus within the adiabatic hyperspherical expansion method
  with realistic nucleon-nucleon interactions. The overlap between the
  $^3$He and the deuteron wave functions can be accurately computed
  from a three-body calculation.  The nucleons are described by
  solutions of the Dirac equation with scalar and vector (S-V)
  potentials. The wave function of the outgoing proton is obtained by
  solving the Dirac equation with a S-V optical potential fitted to
  elastic proton scattering data on the residual nucleus. Within this
  theoretical framework, we compute the cross section of the reaction
  and other observables like the transverse-longitudinal asymmetry,
  and compare them with the available experimental data measured at
  JLab.

\end{abstract}

\section{Introduction and Theoretical Framework}
\label{intro}
Coincidence ($e,e^\prime p$) measurements at quasielastic kinematics 
are a powerful tool to study bound nucleon
properties. Over the years they have provided us with a wealth
of detailed information on bound energies, momentum distributions and 
spectroscopic factors.  This is so because
at quasielastic kinematics the ($e,e^\prime p$) reaction can be
treated with confidence in the Impulse Approximation (IA), {\em i.e.},
assuming that the exchanged photon is absorbed by a single nucleon
which is the one detected. %Where the IA dominates the process, the reaction then is sensitive to single-particle properties of the nucleon. 
Being the electron the probe particle has two important
advantages: experimentally it is easy to get a beam of electrons and
to choose conditions to select independently both the energy and
momentum transferred to the nucleus, and theoretically its interaction
with the nucleus is perfectly described within quantum electrodynamics.

$^3$He is a light nucleus belonging to a simple-to-intermediate
position in the nuclear systems, which can be described within a
Faddeev formalism \cite{nie01}, while the deuteron is the simplest
system of bound nucleons. Hence, this reaction provides an almost
unique case where a theoretical treatment with the least amount of
approximations can be attempted. In this contribution we show 
the results for cross section and transverse-longitudinal asymmetry
($A_{TL}$) for $^3$He($e,e^\prime p$)$^2$H within the Relativistic
Distorted Waves Impulse Approximation (RDWIA), compared to the most
recent experimental data measured at JLab \cite{rva}. These data
have unprecedent accuracy and detail and then constitute a very stringent test of the 
theoretical modelling employed to describe these reactions.

{\raggedleft \em The Relativistic Distorted Wave Impulse Approximation.}
%An electron with energy and momentum $(\varepsilon_i,\bm{k}_i)$ hits a
%target with energy and momentum $(E_I,\bm{P}_I)$. The energy and
%momentum transfered to the nucleus is denoted by $(\omega,\bm{q})$,
%and the energy and momentum of the scattered electron is
%$(\varepsilon_f,\bm{k}_f)$. After the collision one proton with energy
%and momentum $(E_p,\bm{p})$ is knocked out from the target, and
%$(E_F,\bm{P}_F)$ are the energy and momentum of the residual nucleus.
Under the IA assumptions, the cross section is proportional to the
(squared) matrix element given by the sum of the individual current
operators of each of the nucleons in the target.%, {\it i.e.} it is a one-body operator. 
The nuclear current is then given by the expression
\cite{Udias} $
  J^\mu(\omega,\bm{q})=\int d\bm{p} \: 
  \bar {\psi}_F (\bm{p} + \bm{q}) \: \hat{J}^\mu(\omega,\bm{q}) \: \tilde
  \phi(\vec p) \: $
where $| \tilde \phi \rangle = \langle \Psi_{^2H} | \Psi_{^3He} \rangle$ is
the overlap between the initial %($^3$He)
and final %($^2$H)
nuclei wavefunctions, also denoted as the quasiparticle wave function within
this context. $\omega$ and $q$ are the energy and momentum transferred in the 
reaction. %As it is usual 
Within the RDWIA \cite{Udias}, the outgoing proton wave 
function $\psi_F$ is a 4-spinor
solution of the Dirac equation with Scalar and Vector (S-V) optical
potentials fit to describe p-A elastic scattering by the residual system, to
take into account the Final State Interactions (FSI). %To obtain a relativistic expression to the overlap and thus avoiding the non-relativistic reduction of the electromagnetic current operators, from the non-relativistic quasiparticle wave function, a relativistic wave function is built for the quasiparticle with only positive energy projections that (properly normalized) coincides with the non-relativistic overlap.
From the non-relativistic overlap, a relativistic wave function is built
for the quasiparticle with only positive energy projections. The upper
components of this relativistic quasiparticle wavefunction 
(properly normalized) coincide with the non-relativistic overlap.

{\raggedleft \em The Hyperspherical Adiabatic Expansion Method.} The $^3$He
wave function can be obtained by solving the Faddeev equations with
the hyperspherical adiabatic expansion method \cite{nie01}.  We have
considered realistic nucleon-nucleon interactions AV18 \cite{wir95}
and a structureless three-body potential.
%of the type $V_{3b} = S \exp{(-\rho^2/b^2)}$.  
The explicit expression for the wave function can be seen in reference~\cite{nie01}.
%\begin{equation}
%\Psi^{JM} = \frac{1}{\rho^{5/2}}\sum_n f_n (\rho) \Phi_{nJM}
%(\rho,\Omega)\;,
%\end{equation}
%where $\rho$ is the so-called hyperradius related to the Jacobi
%coordinates ($\bm{x,y}$) as $\rho^2 = \bm{x}^2 + \bm{y}^2 $. $\Omega$
%denotes the five angular coordinates \cite{nie01}.
%The overlap can then be
%computed as
%\begin{equation}
%\tilde{\phi}(\bm{y})=\left(\frac{m}{\mu_{12}}\right)^{3/2}
%\int d\bm{x} \Psi_{^2H}^\dag(\bm{x})
%\Psi_{^3He}(\bm{x},\bm{y}) 
%\:,
%\label{over}
%\end{equation}
%in configuration space. %$\tilde \phi (\bm p)$ is obtained as the Fourier transform of $\tilde \phi (\bm y)$.

{\raggedleft \em The Correlated Hyperspherical Harmonic Technique.} The
standard $^3$He wave function obtained by a variational wave function
derived from a realistic Hamiltonian consisting of the AV18
nucleon-nucleon potential and URIX \cite{pud95} three-nucleon
interactions has been also considered \cite{kie93}. Its high accuracy
is well documented and has been used before for similar purposes
\cite{sch05,cio08}. We will consider it as a reference. 
%The overlap is obtained as in eq.~(\ref{over}).

\begin{figure}
\centering
\includegraphics[scale=0.23,angle=-90]{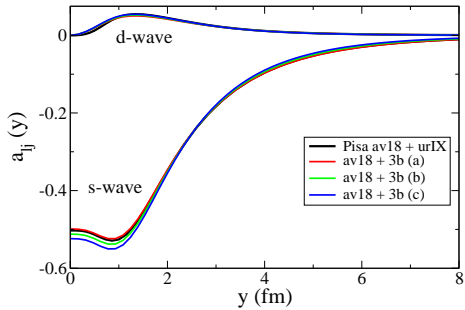}
\caption{Coefficients of the expansion of the overlap into spherical
  harmonics. The black line (labeled as ``Pisa'' corresponds to the
  correlated hyperspherical harmonic technique. The red, green and
  blue lines correspond to the adiabatic calculation. $a$, $b$ and $c$
  refer to three different fits of the three-body energy.}
\label{fig:figureover}
\end{figure}

The overlap %(\ref{over}) 
of the initial and final nuclei wave functions in configuration space can be expanded into spherical harmonics as
$\tilde{\phi}(\bm{y}) = \sum_{lj} a_{lj}(y) \phi _{lj}(\Omega_y)$,
separating the radial part from the angular part.
Fig.~\ref{fig:figureover} shows the expansion coefficients for four
different calculations: one obtained with the variational method from
the correlated hyperspherical harmonic technique (labeled as {\em
  Pisa} in the figure), and the other three have been calculated
within the adiabatic expansion method. If the computations are
accurate enough and well converged, both approximations should be
equivalent. We have considered different 3-body potentials. In the
adiabatic case the three-body potential has been fitted to three
different binding energies: -6.72~MeV (labeled as ``$a$'' in the
figure), -7.72~MeV ($b$) and -8.79~MeV ($c$). The binding energy
corresponding to {\em Pisa} is -7.74~MeV. We can observe that the $a$
curve approaches better the {\em Pisa} one, even though the $b$
binding energy is the experimental one. This has to be attributed to
the different three-body forces employed in this calculation and the
one from the Pisa group. The use of three slightly different overlaps
allows use to test the sensitivity of the measured data to fine
details of the overlap.

\section{Results and Conclusions}

\begin{figure}
\begin{minipage}{0.5\linewidth}
\centering
\includegraphics[scale=0.23,angle=-90]{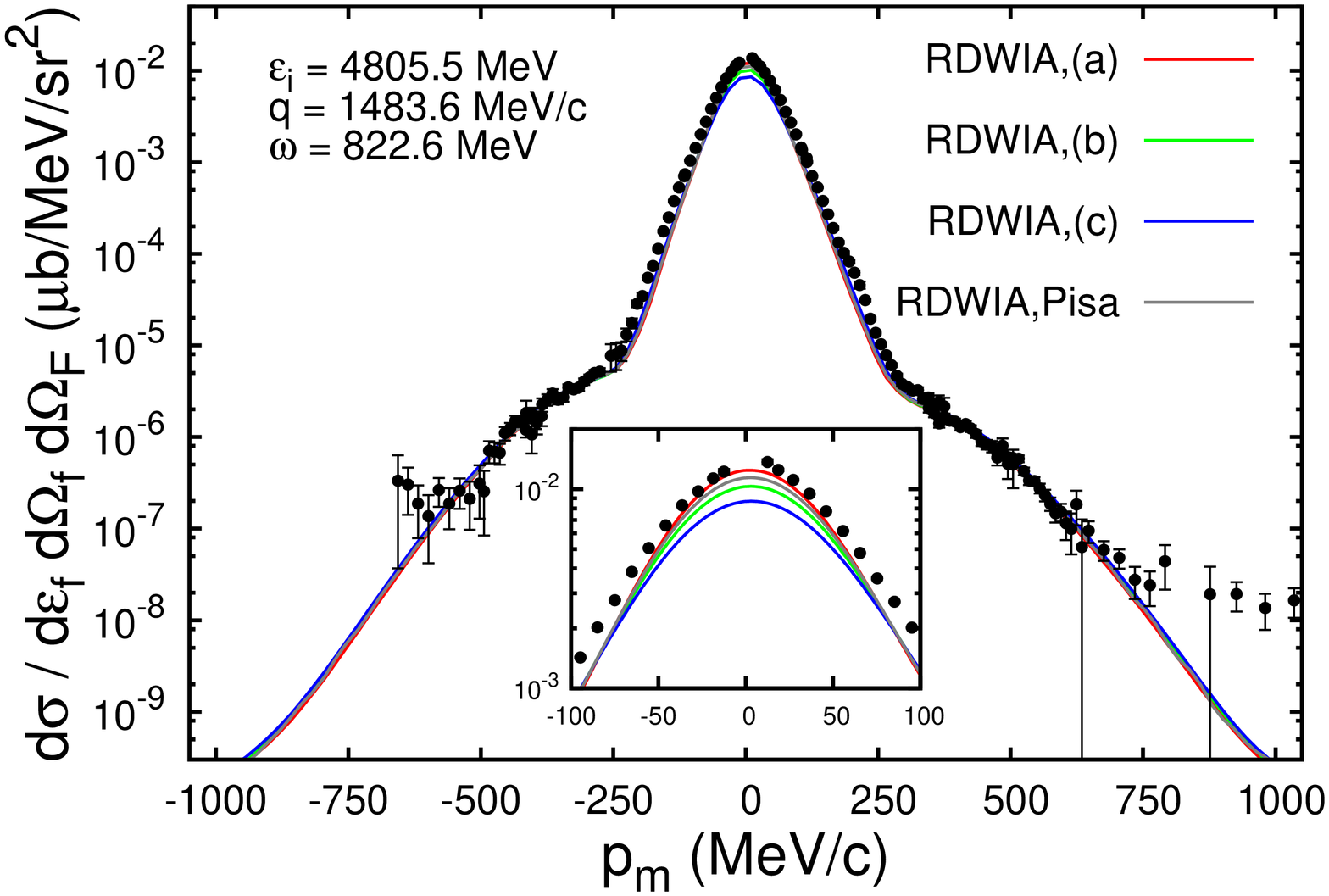}
\caption{The $^3$He($e,e^\prime p$)$^2$H differential cross section as
  a function of the missing momentum. }
\label{fig:figure1}
\end{minipage}
\hspace{0.25cm}
\begin{minipage}{0.5\linewidth}
\centering
\includegraphics[scale=0.23,angle=-90]{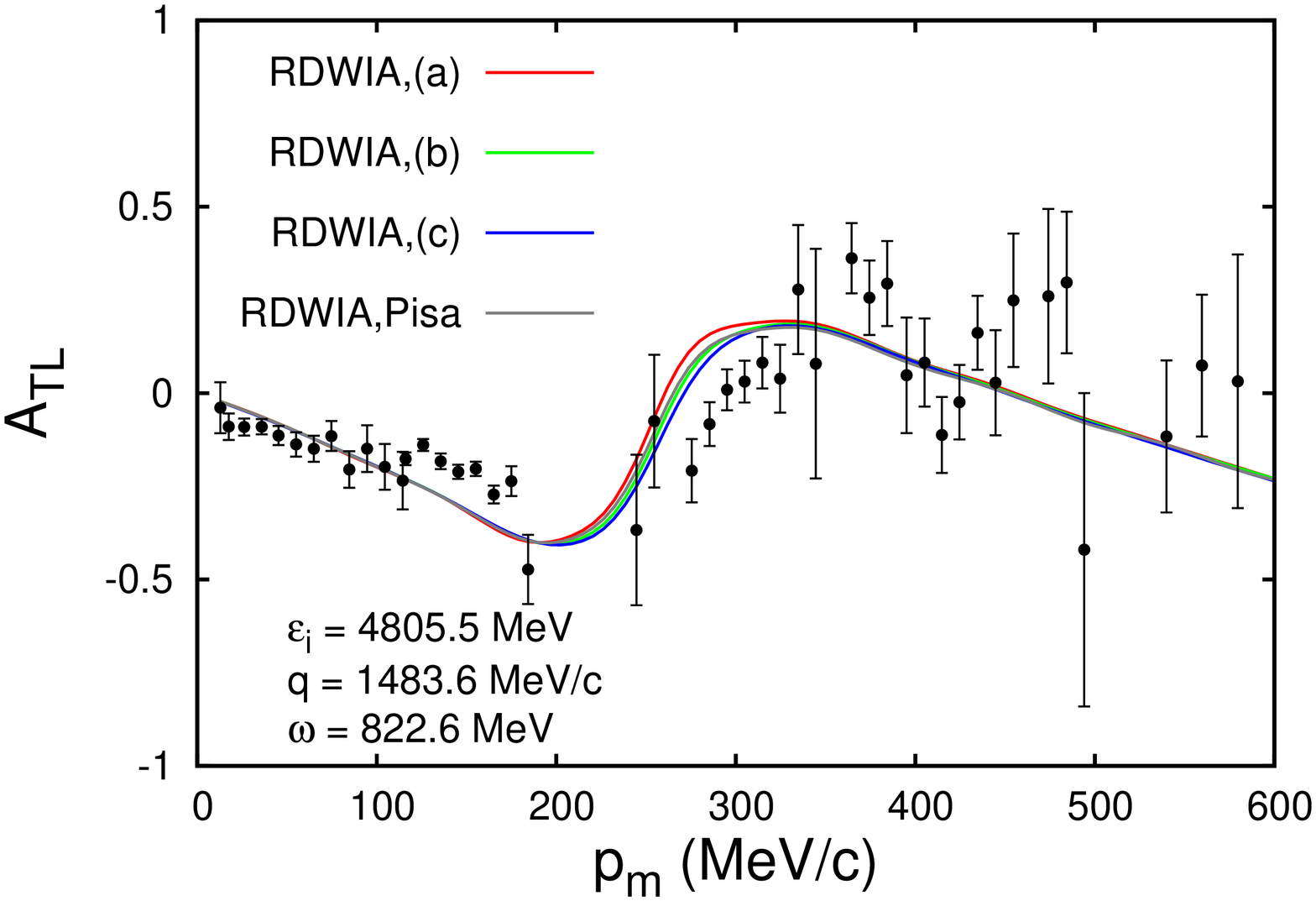}
\caption{The $A_{TL}$ for the $^3$He($e,e^\prime p$)$^2$H reaction as a
  function of the missing momentum.  }
\label{fig:figure2}
\end{minipage}
\end{figure}

Fig.~\ref{fig:figure1} shows the computed cross section of the
reaction $^3$He($e,e^\prime p$)$^2$H together with the data measured
at JLab \cite{rva}. The experimental data are very well reproduced by
our calculations in all the $p_{m}$ range except for the region around
$p_m = 0$. It should be pointed out that the theoretical results are
free from any scale factor. The description of the data both for
cross section and TL asymmetry is very good, even at this level of IA
only contributions and without any adjustable parameter. The agreement
with experiment is comparable or better than the one of
non-relativistic calculations, which is remarkable considering the
difference in the ingredients for the current operator and the final
state, while it is true that the overlap integral is essentially the
same one used in previous analysis \cite{cio08}.%,BGPR96}
In the inset of
fig.~\ref{fig:figure1} it is shown that at low missing momentum the
curve that better approaches the experimental data is the {\em Pisa}
one, followed by the $a$ one. At higher momenta all of the curves are
very similar. This shows that the experimental data are sensitive to
small detail of the overlap and that thus these can be used to refine
aspects of the description of the target nucleus.

The $A_{TL}$ can be obtained as $A_{TL} = \frac{\sigma_+ -
  \sigma_-}{\sigma_+ + \sigma_-} \:,$ where $\sigma_{\pm}$ are the
coplanar cross sections measured at positive and negative missing
momentum. The agreement between theory and experiment is quite
satisfactory, as can only be obtained within a fully unfactorized
calculation as shown by previous work \cite{cio08}.

%\begin{figure*}
%  \includegraphics[width=0.75\textwidth]{example.eps}
%\end{figure*}

To summarize, in this contribution we have studied the
$^3$He($e,e^\prime p$)$^2$H reaction under the RDWIA. The
quasiparticle wave function has been obtained from a three-body
Faddeev calculation with realistic nucleon-nucleon interactions. It
has been compared to the one obtained by the Pisa group. The
calculation is completely parameter free and it is in fair agreement
with the experiment right out of the box, in a pure Impulse
Approximation calculated cross section. We have shown results for the
computed cross section and $A_{TL}$ together with the experimental
data measured at JLab. The good agreement between experimental and
theoretical results found here as well as in previous non-relativistic
approaches that employed different framework to describe FSI is a
strong indication that this reaction is mostly sensitive to the
overlap integral while the other ingredients, once roughly described,
introduce small uncertainty. The good agreement for this
parameter-free case, also vindicates the assumptions usually taken
within RDWIA to analyze $(e,e'p)$ reactions.

\begin{acknowledgements}
  This work was partly supported by funds provided by DGI of MEC
  (Spain) under Contracts No. FPA2007-62216, FIS2008-01301 and the
  Spanish Consolider-Ingenio programme CPAN (Programme
  No. CSD2007-00042). R.A.R. and J.R.V. acknowledge support by
  Ministerio de Ciencia e Innovaci\'on (Spain) under the ``Juan de la
  Cierva'' programme. 
\end{acknowledgements}



\end{document}